\documentclass[12pt]{iopart}
\newcommand{\gguide}{{\it Preparing graphics for IOP journals}}
\begin{document}

\title[Influence of helicity on scaling regimes in ]{Influence of helicity on scaling regimes in the extended Kraichnan model}

\author{O G Chkhetiani$^1$, M Hnatich$^{2,3}$, E Jur\v{c}i\v{s}inov\'a$^{2,4}$,
        M Jur\v{c}i\v{s}in$^{2,5}$, A Mazzino$^{6}$ and M Repa\v{s}an$^{2}$}
\address{$^{1}$ Space Research Institute, Profsoyuznaya 84/32, 117997 Moscow, Russian Federation}
\address{$^{2}$ Institute of Experimental Physics, Slovak Academy of Sciences,
                Watsonova 47, 040 01, Ko\v{s}ice, Slovakia}
\address{$^{3}$  Department of Mathematics Faculty of Civil Engineering, Technical University
Vysoko\v{s}kolsk\'a 4,
040 01 Ko\v{s}ice, Slovakia}
\address{$^{4}$ Laboratory of Information Technologies, Joint Institute for Nuclear Research,
                141980 Dubna, Moscow Region, Russian Federation}
\address{$^{5}$ N.N. Bogoliubov laboratory of Theoretical Physics, Joint Institute for
       Nuclear Research, 141980 Dubna, Moscow Region, Russian Federation}
\address{$^{6}$ Department of Physics, University of Genova, National Institute of Nuclear Physics, Genova Section,
via Dodecaneso 33, I-16146 Genova, Italy}

\ead{hnatic@saske.sk}



\begin{abstract}
{We have investigated  the advection of  a passive scalar quantity
by incompressible helical turbulent flow in the frame of extended Kraichnan model.  Turbulent fluctuations of velocity field
are assumed to  have the Gaussian statistics with zero mean and   defined noise with finite time-correlation.
Actual calculations have been done up to two-loop approximation in the frame of field-theoretic renormalization group approach.
It turned out that space parity violation (helicity) of turbulent environment does not affect anomalous scaling
which is peculiar attribute of corresponding model without helicity. However, stability of asymptotic regimes, where
anomalous scaling takes place, strongly depends on the amount of helicity. Moreover, helicity gives rise to the turbulent
diffusivity, which has been calculated in one-loop approximation.}
\end{abstract}
\pacs{47.27.-i, 47.10.+g, 05.10.Cc}
\submitto{\JPA}

\section{Introduction}

During the last decade much attention has been paid to the
inertial range of fully developed turbulence, which contains wave
numbers larger then those that pump the energy into the system and
smaller enough then those that are related to the dissipation
processes \cite{Monin}. Grounding of the inertial
range turbulence have been created in the well known
 Kolmogorov--Obukhov (KO) phenomenological theory
(see, e.g., \cite{Monin,McComb,Legacy}). One of the main problems in
the modern theory of fully developed turbulence is to verify the
validity of the basic principles of  KO theory and their
consequences
 within the framework of a
microscopic model. Recent experimental and theoretical studies
indicate  possible deviations from the celebrated Kolmogorov
scaling exponents. The scaling behavior  of the velocity
fluctuations with exponents, which values are different from
Kolmogorov ones, is called as anomalous and usually  is associated
with intermittency phenomenon. Roughly speaking, intermittency
means that statistical properties (for example, correlation or
structure functions of the turbulent velocity field) are dominated
by rare spatiotemporal configurations, in which the regions with
strong turbulent activity have exotic (fractal) geometry and are
embedded into the vast regions with regular (laminar) flow. In the
turbulence such phenomenon is believed to be related to the strong
fluctuations of the energy flux which, therefore, leads to
deviations from the predictions of the aforementioned KO theory.
The deviations, referred to as ``anomalous'' or
``non-dimensional'' scaling, manifest themselves in singular
(arguably power-like) dependence of correlation or structure
functions on the distances and the integral (external) turbulence
scale $L$. The corresponding exponents are certain nontrivial and
nonlinear functions of the order of the correlation function, the
phenomenon referred to as ``multiscaling''.

Although the theoretical description of the fluid turbulence on
the basis of the "first principles", i.e., on the stochastic
Navier-Stokes (NS) equation \cite{Monin} remains essentially
an open problem, considerable progress has been achieved in
understanding simplified model systems that share some important
properties with the real problem: shell models \cite{Dyn},
stochastic Burgers equation \cite{Burgulence} and passive
advection by random ``synthetic'' velocity fields \cite{FGV}.

The crucial role in these studies are played by the models of
advected passive scalar field \cite{Obu49}.
A simple model of a passive scalar quantity advected by a random
Gaussian velocity field, white in time and self-similar in space
(the latter property mimics some features of a real turbulent
velocity ensemble), the so-called Kraichnan's rapid-change model
\cite{Kra68}, is an example. The interest to these models is based
on two important facts: first, as were shown by both natural and
numerical experimental investigations, the deviations from the
predictions of the classical Kolmogorov-Obukhov phenomenological
theory \cite{Monin,McComb,Legacy} is even more
strongly displayed for a passively advected scalar field than for
the velocity field itself (see, e.g.,
\cite{AnHoGaAn84,Sre91,HolSig94} and
references cited therein), and second, the problem of passive
advection is much more easier to be consider from theoretical
point of view. There, for the first time, the anomalous scaling
was established on the basis of a microscopic model
\cite{Kraich1}, and corresponding anomalous exponents was
calculated within controlled approximations
\cite{GK,Pumir} (see also review \cite{FGV} and
references therein).

The greatest stimulation to study the simple models of passive
advection not only of scalar fields but also of vector fields
(e.g., weak magnetic field) is related to the fact that even
simplified models with given Gaussian statistics of so-called
"synthetic" velocity field describes a lot of features of
anomalous behavior of genuine turbulent transport of some
quantities (as heat or mass) observed in experiments, see, e.g.,
\cite{HolSig94}--\cite{AveMaj90}
and references cited therein.

The term ``anomalous scaling'' reminds of the critical scaling in
models of equilibrium phase transitions. In those, the field
theoretic methods were successfully employed to establish the
existence of self-similar (scaling) regimes and to construct
regular perturbative calculational schemes (the famous $\epsilon$
expansion and its relatives) for the corresponding exponents,
scaling functions, ratios of amplitudes etc; see e.g.
\cite{Zinn,book}. Here and below, by ``field theoretic methods''
we mean diagrammatic and functional techniques, renormalization
theory and renormalization group, composite operators,
operator-product expansion and so on \cite{book}.

The feature specific to the theory of turbulence and simplified
models associated with it is the existence in the corresponding
field theoretical models of the composite operators with {\it
negative} scaling (critical) dimensions. Such operators, termed
``dangerous'' in \cite{RG,turbo,RG1,Novikov,RG3}, give rise to the
anomalous scaling, i.e., the singular dependence on the infrared
(IR) scale  $L$ with certain nonlinear anomalous exponents.

Important advantages of the RG approach are its universality and
calculational efficiency: a regular systematic perturbation
expansion for the anomalous exponents was constructed, similar to
the well-known $\epsilon$-expansion in the theory of phase
transitions, and the exponents were calculated in the first order
of expansion for passively advected vector fields
\cite{Lanotte2,amodel}  and in the second \cite{RG} and third
\cite{RG1} orders of that expansion for scalar fields.
Furthermore, the RG approach is not related only to the
rapid-change model and can also be applied to the case with finite
correlation time, anisotropy, the space parity violation, and,
moreover,  non-Gaussian advecting field \cite{RG3}.

The solution proceeds in two main stages. In the first stage, the
multiplicative renormalizability of the corresponding field
theoretic model is demonstrated and the differential RG equations
for its correlation functions are obtained. The asymptotic
behavior of the latter on their UV argument $(r/l)$ ($l$ is
internal  length) for $r\gg\ell$ and any fixed $(r/L)$ ($L$  is an
outer length) is given by IR stable fixed points of those
equations. It involves some ``scaling functions'' of the IR
argument $(r/L)$, whose form is not determined by the RG
equations. In the second stage, their behavior at $r\ll L$ is
found from the operator product expansion within the framework
of the general solution of
the RG equations. There, the crucial role is played by the
critical dimensions of various composite operators, which give
rise to an infinite family of independent scaling exponents (and
hence to multiscaling). Of course, these both stages (and thus the
phenomenon of multiscaling) have long been known in the RG theory
of critical behavior. The distinguishing feature, specific to
the models of turbulence  is the existence of composite operators with
aforementioned {\it negative} critical dimensions. Their
contributions to the operator product expansion  diverge at
$(r/L)\to 0$. In the models of critical phenomena, nontrivial
composite operators always have  positive dimensions and
  determine only corrections (vanishing for $(r/L)\to0$)
to the leading terms (finite for $(r/L)\to0$) in the scaling
functions.

Existence of  regular perturbation schemes and accurate numerical
simulations allows one to discuss, for the example of the
rapid-change model and its descendants, the issues that are
interesting within the general context of fully developed
turbulence: universality and saturation of anomalous exponents,
effects of compressibility, anisotropy and pressure, persistence
of the large-scale anisotropy and hierarchy of anisotropic
contributions and so on. Moreover, it is interesting and important
to study  the helicity (violation of space parity) effects because
many turbulence phenomena are directly influenced by them.

In \cite{RG3} the problem of a passive scalar advected by the
Gaussian self-similar velocity field with finite correlation time
\cite{all2} was studied by the field theoretic RG method. There, the
systematic study of the possible scaling regimes and anomalous
behavior was present at one-loop level. The two-loop corrections to
the anomalous exponents were obtained in \cite{Juha2}. It was shown
that the anomalous exponents are nonuniversal as a result of their
dependence on a dimensionless parameter, the ratio of the velocity
correlation time, and  turnover time of a scalar field.

In what follows, we shall continue with the investigation of this
model from the point of view of the influence of helicity on the
scaling regimes and the anomalous exponents within two-loop
approximation.

Helicity is defined as the scalar product of velocity and
vorticity and its non zero value expresses mirror symmetry
breaking of turbulent flow. It plays significant role in the
processes of magnetic field generation in electrically conductive
fluid \cite{dynamo1,Moffatt} and represents one of the most
important characteristics of large-scale motions as well
\cite{Etling85,Ponomar2003}.
Despite of this fact
the role of the helicity in hydrodynamical turbulence is not
completely clarified up to now.

The Navier-Stokes equations conserve kinetic energy and helicity
in inviscid limit. Presence of two quadratic invariants leads to
the possibility of appearance of double cascade. It means that
cascades of energy and helicity take place in different ranges of
wave numbers analogously to the two-dimensional turbulence and/or
the helicity cascade  appears concurrently to the energy one in
the direction of small scales \cite{Brissaud73,Moiseev1996,koprovall}.
Particularly,  helicity cascade is closely connected with the
existence of exact relation between triple and double correlations
of velocity known as ``2/15'' law analogously to the ``4/5''
Kolmogorov law \cite{Chkhet96,kurien}. Corresponding to \cite{Brissaud73}
aforementioned scenarios of turbulent cascades  differ each other
by spectral scaling. Theoretical arguments given by Kraichnan
\cite{Kraich73}  and results of numerical calculations  of
Navier-Stokes equations \cite{Chen2003} support
the scenario of concurrent cascades. The appearance of helicity in
turbulent system leads to constraint of non-linear cascade to the
small scales. This phenomenon was firstly demonstrated by
Kraichnan \cite{Kraich73} within the modelling problem of
statistically equilibrium spectra and later in numerical
experiments.

Turbulent viscosity and diffusivity, which characterize influence of
small-scale motions on heat and momentum transport, are basic
quantities investigated in  the theoretic and applied models. The
constraint of direct energy cascade in helical turbulence  has to be
accompanied by decrease of turbulent viscosity. However, no
influence of helicity on turbulent viscosity was found in some works
\cite{Pouquet78,Zhou91}. Similar situation is observed for turbulent
diffusivity in helical turbulence. Although the modelling
calculations demonstrate intensification of turbulent transfer in
the presence of helicity \cite{Drum84} direct calculation of
diffusivity  does not confirm this effect
\cite{Knobl77,Lipscombe91}. Helicity is the pseudoscalar quantity
hence it can be easily understood, that its influence appears only
in quadratic and higher terms of perturbation theory or in the
combination with another pseudoscalar quantities (e.g., large-scale
helicity). Really, simultaneous consideration of memory effects and
second order approximation indicate effective influence of helicity
on turbulent viscosity \cite{Belian1998} and turbulent diffusivity
\cite{Drum84,Dolg87,Drummond2001} already in the limit of small or
infinite correlation time.

Helicity, as we shall see below, does not affect known results in
one-loop approximation and,  therefore, it is necessary to turn to
the second order (two-loop) approximation to be able to analyze
possible consequences. It is also important to say that in the
framework of classical Kraichnan model, i.e., the model of passive
advection by the Gaussian velocity field with $\delta$-like
correlations in time, it is not possible to study the influence of
the helicity because all potentially "helical" diagrams are
identically equal to zero at all orders in the perturbation
theory. In this sense, the investigation of the helicity in the
present model can be consider as the first step to analyze the
helicity in genuine turbulence.

\section{Field theoretic description of the model}

The advection of a passive scalar field
$\theta(x)\equiv\theta(t,\mathbf{x})$ in helical turbulent
environment is described by the stochastic equation
\begin{equation}
\partial_t \theta + v_i \partial_i \theta=\nu_0 \Delta
\theta + f,\label{eq:theta}
\end{equation}
\noindent where $\partial_{t}\equiv\partial/\partial t$,
$\partial_{i}\equiv\partial/\partial \mathrm{x}_{i}$, $\nu_{0}$ is the
molecular diffusivity coefficient (hereafter all parameters with a
subscript $0$ denote bare parameters of unrenormalized theory; see
below), $\triangle\equiv\partial^{2}$ is the Laplace operator, $v_i
\equiv v_i(x)$ is the $i$-th component of the divergence-free (owing
to the incompressibility) velocity field ${\bf v}(x)$, and $f\equiv
f(x)$ is an artificial the Gaussian random noise with zero mean and
correlation function
\begin{equation}
\langle f(x)f(x')\rangle  =  \delta(t-t')C(\mathbf{r}/L),\,\,\,
\mathbf{r} = \mathbf{x}-\mathbf{x}',\label{eq:corelf}
\end{equation}
where $L$ denotes an integral (outer) scale. It maintains the steady-state
of the system but the detailed form of the function
$C(\mathbf{r}/L)$
is unessential in our
consideration.
In
spite of the fact that in real problems the velocity field ${\bf
v}(x)$ satisfies the Navier-Stokes equation, in what follows, we suppose
that the statistics of velocity field is given in the form of Gaussian
distribution with zero mean and correlator
\begin{equation}
\langle v_i(x) v_j(x^{\prime}) \rangle  =  \int \frac{\mathrm{d}\omega d^d
k}{(2\pi)^{d+1}} P^{\rho}_{ij}({\bf k}) D_v(\omega,k)
\exp[-i\omega(t-t^{\prime})+i{\bf k}({\bf x}-{\bf x^{\prime}})],
\label{eq:corelv}
\end{equation}
with
\begin{equation}
D_v(\omega, k) = \frac{D_0
k^{4-d-2\varepsilon-\eta}}{(i\omega+u_0 \nu_0
{k}^{2-\eta})(-i\omega+u_0 \nu_0 k^{2-\eta})},\label{corrvelo}
\end{equation}
where $k=|\mathbf{k}|$,  $D_0=\mathrm{g}_0\nu_0^3$ is a positive amplitude factor,  $\mathrm{g}_0$  plays the role of
the coupling constant of the model, an  analog of the coupling
constant $\lambda_0$ in the $\lambda_0 \varphi^4$ model of critical
behavior \cite{Zinn,book}. In addition, $\mathrm{g}_0$ is a formal
small parameter of the ordinary perturbation theory. The positive
exponents $\varepsilon$ and $\eta$ ($\varepsilon=O(\eta)$) are small
RG expansion parameters, the analogs of the parameter
$\varepsilon=4-d$ in the $\lambda_0 \varphi^4$ theory. Thus we have
a kind of double expansion model in the $\varepsilon -\eta$ plane
around the origin $\varepsilon=\eta=0$. The correlator
(\ref{corrvelo}) is directly related to the energy spectrum via the
frequency integral \cite{RG3}
\begin{equation}
E(k)\simeq k^{d-1} \int \mathrm{d}\omega D^v(\omega, k) \simeq
\frac{\mathrm{g}_0 \nu_0^2}{u_0} k^{1-2\varepsilon}.
\end{equation}
Therefore, the coupling constant $\mathrm{g}_0$ and the exponent
$\varepsilon$ describe the equal-time velocity correlator or,
equivalently, energy spectrum. On the other hand, the constant $u_0$
and the second exponent $\eta$ are related to the frequency $\omega
\simeq u_0 \nu_0 k^{2-\eta}$  which characterizes the mode $\mathbf{k}$
\cite{Eyink96}. Thus, in our notation, the
value $\varepsilon=4/3$ corresponds to the well-known Kolmogorov
"five-thirds law" for the spatial statistics of velocity field, and
$\eta=4/3$ corresponds to the Kolmogorov frequency.
For completeness, we remain $d$-dependence in expressions
(\ref{eq:corelv}) and (\ref{corrvelo}) ($d$ is the dimensionality of
the ${\bf x}$ space), although, of course, when one investigates
system with helicity the dimension of the ${\bf x}$ space must be
strictly equal to three.
To include helicity the transverse projector $P^{\rho}_{ij}({\bf k})$ is
taken in the form
\begin{equation}
P^{\rho}_{ij}({\bf k})=P_{ij}({\bf k}) + H_{ij}({\bf k})=
\delta_{ij}-k_{i}k_{j}/k^{2}+i\rho\epsilon_{ijl}\frac{k_{l}}{k}.\label{eq:Pij}
\end{equation}
Here $P_{ij}({\bf k})=\delta_{ij}-k_{i}k_{j}/k^{2}$ represents
non-helical part of the total transverse projector
$P^{\rho}_{ij}({\bf k})$. On the other hand, $H_{ij}({\bf
k})=i\rho\epsilon_{ijl}k_{l}/k$ mimics the presence of helicity in
the flow. Thus, formally, the transition to the helical fluid
corresponds to the breaking  of spatial parity, and, technically,
this is expressed by the fact that the correlation function is
specified in the form of mixture of a true tensor and a
pseudotensor. The tensor $\varepsilon_{ijl}$ is Levi-Civita's
completely antisymmetric tensor of rank 3 and the real parameter
$\rho$, characterizes the amount of helicity. Due to the requirement
of positive definiteness of the correlation function the absolute
value of $\rho$ must be in the interval $|\rho| \in \langle
0,1\rangle$. Non-zero helical part proportional to $\rho$ physically
expresses existence of non-zero correlations $\langle{\bf v}\cdot
\mathrm{rot}\:{\bf v}\rangle$.

The general model (\ref{eq:corelv}), (\ref{corrvelo}) contains two
important special cases: rapid-change model limit when
$u_0\rightarrow \infty$ and $\mathrm{g}_0^{\prime}\equiv
\mathrm{g}_0/u_0^2=$ const, $D_v(\omega, k)\rightarrow
\mathrm{g}_0^{\prime} \nu_0 k^{-d-2\varepsilon + \eta},$ and
quenched (time-independent or frozen) velocity field limit which is
defined by $u_0\rightarrow 0$ and $\mathrm{g}_0^{\prime\prime}\equiv
\mathrm{g}_0/u_0=$ const, $D_v(\omega, k)\rightarrow
\mathrm{g}_0^{\prime\prime} \nu_0^2\pi \delta(\omega)
k^{-d+2-2\varepsilon},$ which is similar to the well-known models of
the random walks in random environment with long range correlations;
see, e.g., \cite{Bouchaud,Honkonen}.

Using Martin-Siggia-Rose mechanism \cite{Martin} the stochastic
problem (\ref{eq:theta})-(\ref{corrvelo}) can be treated as a
field theory with action functional
\begin{equation}
S(\theta,\theta',\mathbf{v})=\theta'D_{\theta}\theta'/2+
\theta'[-\partial_{t}+\nu_{0}\triangle-(v_i\partial_i)]\theta-
\mathbf{v}D_{v}^{-1}\mathbf{v}/2,\label{eq:Sucinok}
\end{equation}
where $\theta'$ is an auxiliary scalar field, and $D_{\theta}$ and
$D_{v}$ are correlators (\ref{eq:corelf}) and (\ref{eq:corelv}),
respectively. In the action (\ref{eq:Sucinok}) all the required
integrations over $x=(t,\mathbf{x})$ and summations over the
vector indices are understood. The first four terms in
Eq.\,(\ref{eq:Sucinok}) represent the Dominicis-Jansen type
action for the stochastic problem (\ref{eq:theta}),
(\ref{eq:corelf}) at fixed $\mathbf{v}$, and the last term
represents the Gaussian averaging over $\mathbf{v}$.

\begin{figure}[t]
     \vspace{0.0cm}
    \begin{center}
       \leavevmode
      \epsfxsize=10cm
     \epsffile{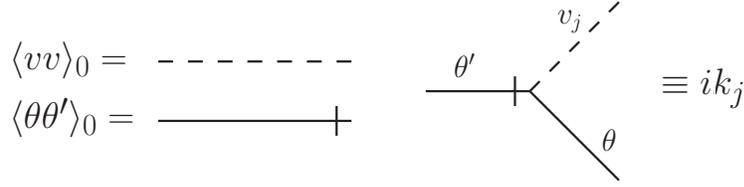}
\end{center}
\vspace{0.0cm} \caption{(Left) Graphical representations of the
needed propagators of the model. (Right) The triple (interaction)
vertex of the model. Momentum ${\mathbf k}$ is entering into the vertex
via field $\theta^{\prime}$. \label{fig1}}
\end{figure}

The model (\ref{eq:Sucinok}) corresponds to a standard Feynman
diagrammatic technique with the  bare propagators $\langle
\theta\theta^{\prime} \rangle_0$ and $\langle v_i v_j\rangle_0$
(in the momentum-frequency representation)
\begin{equation}
\langle \theta(\omega, {\mathbf{k}})\theta^{\prime}(\omega, {\mathbf{k}})
\rangle_0 = \frac{1}{-i\omega+\nu_0
k^2}, \,
\langle v_i(\omega, {\mathbf{k}}) v_j(\omega, {\mathbf{k}})\rangle_0 =
P^{\rho}_{ij}(\mathbf{k}) D_v(\omega, k), \label{proptheta}
\end{equation}
where $D_v(\omega, k)$ is given directly by (\ref{corrvelo}). In the
Feynman diagrams, these propagators are represented by the lines
which are shown in figure \ref{fig1} (the end with a slash in the
propagator $\langle \theta\theta^{\prime} \rangle_0$ corresponds to
the field $\theta^{\prime}$, and the end without a slash corresponds
to the field $\theta$). The triple vertex (or interaction vertex)
$-\theta^{\prime} v_j\partial_j \theta = \theta^{\prime} v_j V_j
\theta $, where $V_j=i k_j$ (in the momentum-frequency
representation), is presented in figure \ref{fig1}, where momentum
$\mathbf{k}$ is flowing into the vertex via the auxiliary field
$\theta^{\prime}$.

\input epsf

\section{Renormalization group analysis}

The model (\ref{eq:Sucinok}) is logarithmic for $\epsilon=\eta=0$
(the coupling constant $\mathrm{g}_{0}$ is dimensionless) and, in this case,
possible ultraviolet (UV) divergences have the form of poles in
various linear combinations of $\varepsilon$ and $\eta$ in the
correlation functions. Using the standard analysis of quantum field
theory one finds that  all
divergences  can be removed by the only counterterm of the form
$\theta' \triangle \theta$ \cite{RG3}. Thus, the model is
multiplicatively renormalizable, which is expressed explicitly in
the multiplicative renormalization of the parameters $\mathrm{g}_0, u_0$, and
$\nu_0$ in the form
\begin{equation}
\nu_0=\nu Z_{\nu},\,\,\, \mathrm{g}_0=\mathrm{g} \mu^{2\varepsilon+\eta}
Z_{\mathrm{g}},\,\,\,u_0=u\mu^{\eta} Z_u. \label{zetka}
\end{equation}

Here the dimensionless parameters $\mathrm{g},u$ and $\nu$ are the
renormalized counterparts of the corresponding bare ones, $\mu$ is
the renormalization mass (a scale setting parameter),  an artefact
of dimensional regularization. Newly introduced quantities
$Z_i=Z_i(\mathrm{g},u;d,\rho;\varepsilon,\eta)=Z_i(\mathrm{g},u;d,\rho;\varepsilon)$ are
renormalization constants (note that if $\rho$ is non-zero then
$d=3$) and, in general, contain poles of linear combinations of
$\varepsilon$ and $\eta.$ However, as detailed analysis shows,  to
obtain all important quantities as the $\gamma$-functions,
$\beta$-functions, coordinates of fixed points, and the critical
dimensions, the knowledge of the renormalization constants  for
the special choice $\eta=0$ is sufficient up to two-loop
approximation (see details in \cite{RG3}).

The renormalized action functional has the following form
\begin{equation}
S_{R}(\theta,\theta',\mathbf{v})=\theta'D_{\theta}\theta'/2+\theta'[-\partial_{t}+\nu
Z_1\triangle-(v\partial)]\theta-
\mathbf{v}D_{v}^{-1}\mathbf{v}/2 ,\label{eq:Srenorm}
\end{equation}
where the correlator $D_v$ is written in renormalized parameters.
By comparison of the renormalized action (\ref{eq:Srenorm}) with
definitions of the renormalization constants $Z_i$, $i=\mathrm{g},u,\nu$
(\ref{zetka}) one comes to the relations among  them:
\begin{equation}
Z_{\nu}=Z_1,\,\,\, Z_{\mathrm{g}}=Z_{\nu}^{-3},\,\,\,
Z_u=Z_{\nu}^{-1}.\label{zetka1}
\end{equation}
The second and third relations are consequences of the absence of
the renormalization of the term with $D_v$ in renormalized action
(\ref{eq:Srenorm}). The parameter $\rho$ a the fields $\theta, \theta', \mathbf{v}$ are not renormalized therefore $Z_{\rho}=Z_{\theta}=Z_{\theta'}=Z_{\mathbf{v}}=1$.

The issue of interest is, in particular, the behavior of the
equal-time structure functions
\begin{equation}
S_{n}(r)\equiv\langle[\theta(t,{\bf x})-\theta(t,{\bf x'})]^{n}\rangle,
\qquad \, r\equiv |\mathbf{r}|=|\mathbf{x}-\mathbf{x}'|
\label{struc}
\end{equation}
in the inertial range, specified by the inequalities
$l <<r<<L$ ($l$ is internal length).
Here parentheses $< >$ mean the functional average over the fields $\theta, \theta', {\bf v}$
with the weight $exp(S_R).$ In the isotropic case, the
odd functions $S_{2n+1}$ vanish, while for $S_{2n}$ simple
dimensionality considerations give
\begin{equation}
S_{2n}(r)=  \nu_0^{-n}\, r^{2n}\, R_{2n} ( r/l, r/L, \mathrm{g}_0, u_0, \rho),
\label{strucdim}
\end{equation}
where  $R_{2n}$ are some functions of dimensionless variables. In
principle, they can be calculated within the ordinary perturbation
theory (i.e., as series in $\mathrm{g}_{0}$), but this is not useful
for studying inertial-range behavior: the coefficients are singular
in the limits $\ r/l \to\infty$  and/or $r/L\to 0$, which compensate
the smallness of $\mathrm{g}_{0}$, and in order to find correct
infrared behavior we have to sum the entire series. The desired
summation can be accomplished using the field theoretic
renormalization group (RG) and operator product expansion (OPE); see
\cite{RG,RG1,RG3} for details.

The RG analysis consists of two main stages. On the
first stage, the multiplicative renormalizability of the model is
demonstrated and the differential RG equations for its correlation
(structure) functions  are obtained. The asymptotic behavior of the functions
like (\ref{struc}) for $ r/l>>1$ and any fixed $r/L$ is given
by IR stable fixed points $\mathrm{g}_*, u_*$ (see below) of the RG equations and has the form
\begin{equation}
S_{2n}(r)= \nu_0^{-n}\, r^{2n}\,
(r/l)^{-\gamma_{n}} R_{2n} (r/L, \rho),  \quad r/l>>1
\label{strucdim2}
\end{equation}
with certain, as yet unknown, scaling functions $R_{2n} (r/L, \rho)\equiv R_{2n} (1, r/L, \mathrm{g}_*, u_*, \rho)$.
In the theory of critical phenomena \cite{Zinn,book} the quantity
$\Delta[S_{2n}]\equiv-2n + \gamma_{n}$ is termed the critical
dimension,
and the exponent $\gamma_{n}$, the difference between the critical
dimension $\Delta[S_{2n}]$ and the canonical dimension $-2n$,
is called the anomalous dimension. 

On the second stage, the small $r/L$ behavior of the functions
$ R_{2n} (r/L, \rho)$ is studied within the general representation
(\ref{strucdim2}) using the operator product expansion (OPE).
It shows that, in the limit $r/L\to 0$,
the functions $ R_{2n} (r/L,\rho)$ have the asymptotic forms
\begin{equation}
 R_{2n} (r/L) = \sum_{F} C_{F}(r/L)\, (r/L)^{\Delta_n},
\label{ope}
\end{equation}
where $C_{F}$ are coefficients regular in $r/L$. In general,
the summation is implied over certain  renormalized composite
operators $F$  with critical dimensions $\Delta_n$.
In case under consideration the leading operators $F$ have the form
$F_n= (\nabla_i \theta \nabla_i \theta)^n. $

We have performed the complete two-loop calculation of
the critical dimensions  of the composite
operators $F_n$ for arbitrary values of $n$, $d$, $u$ and $\rho$ and obtain them
in the following form:
\begin{equation}
\Delta_n=\Delta_{n}^{(1)}\epsilon + \Delta_{n}^{(2)}\epsilon^2\, , \qquad \Delta_{n}^{(1)} = \frac{-n(n-2)(d-1)} {2(d-1)(d+2)}
\label{a00}
\end{equation}
where
$\Delta_{n}^{(1)}$ is critical dimension  obtained in one-loop approximation.
Interesting technical details of these two-loop calculations will be present elsewhere.

Two-loop contribution $\Delta_{n}^{(2)}$ is rather cumbersome and
can be found in \cite{Juha2}.  The main and interesting result
consists in the fact that although separated two-loop Feynman graphs
of operators $F_n$ strongly depend on helicity parameter $\rho$,
such dependence disappears in their sum, which gives rise to the
critical  dimensions $\Delta_n$. We can conclude that in two-loop
approximation  anomalous scaling with {\it negative} exponents
(\ref{a00}) is not affected by the existence of non-zero helical
correlations $\langle{\bf v} rot {\bf v}\rangle$ in turbulent
incompressible flow. It turns out, however, that region of stability
of possible asymptotic  regimes governed by fixed points of RG
equations,
 where anomalous scaling
takes place,   and  effective
diffusivity  strongly depend on $\rho$.

\begin{figure}[t]
     \vspace{0.0cm}
\begin{center}
       \leavevmode
       \epsfxsize=12cm
       \epsffile{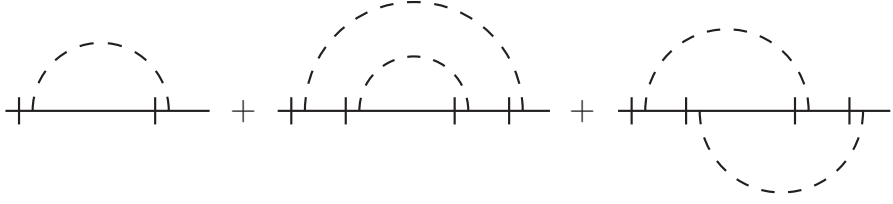}
\end{center}
\vspace{0.0cm} \caption{The one and two-loops contributions to the
self-energy operator $\Sigma$.
\label{fig2}}
\end{figure}

Let us analyze asymptotic regimes in detail. The structure functions
and the other statistical averages of random fields $\theta,
\theta'$ satisfy linear differential RG equations with linear
differential operator ${\cal{D}}_{RG}:$
\begin{equation}
{\cal{D}}_{RG} = {\cal{D}}_{\mu} +
\beta_{\mathrm{g}}(\mathrm{g},u)\partial_{\mathrm{g}}+\beta_u(\mathrm{g},u)\partial_u-\gamma_{\nu}(\mathrm{g},u){\cal{D}}_{\nu}\,
. \label{RGoper}
\end{equation}
Here ${\cal{D}}_x\equiv x\partial_x$ stands for any variable $x$ and
the RG functions (the $\beta$ and $\gamma$ functions) are given by
well-known definitions and in our case, using the relations
(\ref{zetka1}) for the renormalization constants, they acquire the
following form
\begin{eqnarray}
\gamma_{\nu}&\equiv& \tilde{\cal{D}}_{\mu} \ln Z_{\nu}, \label{gammanu}\\
\beta_\mathrm{g}&\equiv&\tilde{\cal{D}}_{\mu} \mathrm{g} =\mathrm{g}
(-2\varepsilon-\eta+3\gamma_{\nu}), \label{betag}\\
\beta_u&\equiv&\tilde{\cal{D}}_{\mu} u =u
(-\eta+\gamma_{\nu}).\label{betau}
\end{eqnarray}
The renormalization constant $Z_{\nu}$ is determined by the
requirement that
response function $G\equiv\langle \theta \theta'\rangle$ must be UV
finite when is written in renormalized variables. In our case it
means that it  has no singularities in the limit $\varepsilon,
\eta\rightarrow 0$. The response function $G$ is related to the
self-energy operator $\Sigma$, which is
expressed via Feynman graphs,  by the Dyson equation. In frequency-momentum
representation it has the following form
\begin{equation}
G(\omega, \mathbf{p})=\frac{1}{-i\omega+\nu_0 p^2 -
\Sigma(\omega, \mathrm{p})}.\label{Dyson}
\end{equation}
Thus, $Z_{\nu}$ is found from the requirement that the UV
divergences are canceled in (\ref{Dyson}) after substitution
$\nu_0=\nu Z_{\nu}$. This determines $Z_{\nu}$ up to an UV finite
contribution, which is fixed by the choice of the renormalization
scheme. In the MS scheme all the renormalization constants have the
form: 1 + {\it poles in $\varepsilon,\eta$ and their linear
combinations}. In contrast to the rapid-change model, where only
one-loop diagram exists (it is related to the fact that all
higher-order loop diagrams contain at least one closed loop which is
built on by only retarded propagators, thus are automatically equal
to zero), in the case with finite correlations in time of the
velocity field, higher-order corrections are non-zero. In two-loop
approximation the self-energy operator $\Sigma$ is defined by
diagrams which are shown in figure \ref{fig2}.

\input epsf

As was already mentioned, in our calculations we can put $\eta=0$.
This possibility essentially simplifies the evaluations of all
quantities \cite{RG3,Juha2}.

 Two-loop calculations of divergent parts of diagrams in figure \ref{fig2} give the renormalization
 constant $Z_{\nu}$ and anomalous dimension $\gamma_{\nu}$ (\ref{gammanu}) in the form:

\begin{equation}
Z_{\nu} = 1 + \frac{g}{\varepsilon} {\cal A} +\frac{g^2}{\varepsilon} {\cal B}+ \frac{g^2}{\varepsilon^2} {\cal C}
\,, \, \gamma_{\nu}=-2(g{\cal A} + 2 g^2 {\cal B}).
\label{nove}
\end{equation}
Here
${\cal A}= (1-d)/(2 d(1+u))$
is the one-loop contribution to the constant $Z_{\nu}$ and anomalous dimension $\gamma_{\nu}$
and the two-loop ones are
\begin{eqnarray}
{\cal B} &=& \frac{(d-1)(d+u)}{4 d^2(d+2)(1+u)^5}
\,\cdot\,{_2F_1}\left(1,1;2+\frac{d}{2};\frac{1}{(1+u)^2}\right) \nonumber
\\ & &
- \frac{\pi \rho^2}{36 (1+u)^3}
\,\cdot\,{_2F_1}\left(\frac12,\frac12;\frac52;\frac{1}{(1+u)^2}\right),\label{BB}\\
{\cal C}& = & -\frac{(d-1)^2}{d^2}\frac{1}{8(1+u)^3}\,,\nonumber
\end{eqnarray}
where
${_2F_1}(a,b,c,z)=1+\frac{a\,
b}{c\cdot1}z+\frac{a(a+1)b(b+1)}{c(c+1)\cdot1\cdot2}z^{2}+\ldots$
represents the hypergeometric function.
We substitute $d=3$ in the helical part (proportional to the $\rho$),
but for completeness we remain the $d$-dependence in the non-helical one.
In addition   we have introduced new notation $g =\mathrm{g} S_d/(2u(2\pi)^d) $
($S_d=2 \pi^{d/2}/\Gamma(d/2)$ denotes the $d$-dimensional
sphere).

From the expressions (\ref{gammanu}) - (\ref{betau}) and
(\ref{nove}) we are able to find and clasify all fixed points $g_*,
u_*$ which satisfy equations:
\begin{equation}
\beta_g(g_*,u_*)=\beta_u(g_*,u_*)=0.
\end{equation}
To investigate the infrared   stability of
a fixed point it is enough to analyze the eigenvalues of the  $2\times2$ matrix
$\Omega$ of first derivatives: $\Omega_{ij}=\partial \beta_{g_i}/\partial g_j$ $(g_i\equiv g, u).$
The anomalous scaling is governed by the infrared stable fixed
points, i.e., those for which both eigenvalues are positive.

Classification and detailed analysis of all fixed points, determination of region of their stability and influence of helicity
will be present elsewhere. Here we confine ourselves to the most interesting  IR stable fixed point, where both parameters $g_*, u_*$ acquire
non-trivial values at $\eta=\varepsilon$:
\begin{eqnarray}
g_*&=& \left((g_*^{(1)}+\left(g_*^{(2)}+g_*^{(3)}\rho^2\right)\varepsilon\right)\varepsilon,\quad
g_*^{(1)}= \frac{3}{2} (1+u_*)\,,
\nonumber \\
 g_*^{(2)}&=& \frac{3 (3+u_*)}{20(1+u_*)^2}\,\cdot\,
{_2F_1}\left(1,1;\frac{7}{2};\frac{1}{(1+u_*)^2}\right)\,,
\label{fix} \\
 g_*^{(3)}&=& - \frac{3 \pi}{8}\,\cdot\,
{_2F_1}\left(\frac12,\frac12;\frac{5}{2};\frac{1}{(1+u_*)^2}\right).
\end{eqnarray}
 Actually, the equation
(\ref{fix}) represents a line of fixed points in $g-u$ plane.
The competition between helical and non-helical terms appears
which  yields a nontrivial restriction for the fixed point values of
variable $u$ to have positive fixed values for variable $g$.

\section{Effective diffusivity \label{sec:EffDiff}}

One of the interesting object from the theoretical as well as
experimental point of view is so-called effective diffusivity $\bar
\nu$. In this section let us briefly investigate the effective
diffusivity $\bar \nu$, which replaces initial molecular diffusivity
$\nu_0$ in (\ref{eq:theta}) due to the interaction of the scalar
field $\theta$ with random velocity field $\mathbf{v}$. Molecular
diffusivity $\nu_0$ governs exponential dumping in time all
fluctuations in the system in the lowest approximation, which is
given by the propagator (response function) (\ref{proptheta}).
Analogously, the effective diffusivity $\bar \nu$ governs
exponential dumping of all fluctuations described by full response
function, which is defined by Dyson equation (\ref{Dyson}). Its
explicit expression can be obtained by the RG approach. In
accordance with general rules of the RG (see, e.g., \cite{book}) all
principal parameters of the model $\mathrm{g}_0, u_0$ and $\nu_0$
are replaced by their effective (running) counterparts, which
satisfy  RG flow equations
\begin{equation}
s \frac{d \bar g}{d s}=\beta_{\mathrm{g}}(\bar g,\bar u)\,,\qquad
s \frac{d \bar u}{d s}=\beta_{u}(\bar g, \bar u)\qquad
s \frac{d \bar \nu}{d s}= -\bar \nu \gamma_{\nu}(\bar g,\bar u)
\label{nu}
\end{equation}
with initial conditions $\bar g|_{s=1}=g, \bar u|_{s=1}=u, \bar
\nu|_{s=1}=\nu$. Here $s=k/\mu$,  $\beta$ and $\gamma$ functions are
defined in (\ref{gammanu})--(\ref{betau}) and  the running
parameters $\bar{g}, \bar{u}$, and $\bar{\nu}$ clearly depend on
variable $s$.
Due to special form of $\beta$-functions (\ref{betag}), (\ref{betau}) we are
able to solve the last equation (\ref{nu}) analytically.
Using the first equation (\ref{nu}) and (\ref{betag}) one immediately rewrites the equation for effective diffusivity
in the form
\begin{equation}
\frac{d\bar  \nu}{\bar \nu}= \frac{\gamma_{\nu}}{2\varepsilon+\eta - 3 \gamma_{\nu}}\frac{d \bar g}{\bar g}
\label{nu1}
\end{equation}
which can be easily integrated. Using initial conditions the
solution acquires the form:
\begin{equation}
\bar \nu = (\frac{g\nu^3}{\bar g s^{2\epsilon + \eta}})^{1/3}=
(\frac{D_0}{\bar g k^{2\epsilon + \eta}})^{1/3}\, .
\label{nue}
\end{equation}
We
emphasize that above solution is exact, i.e., the exponent
$2\epsilon+\eta$ is exact too. However, in infrared region $k\ll
\mu\sim l^{-1}$,  $\bar g \rightarrow g_*,$ which can be calculated
only pertubatively. In the two-loop approximation $g_*=
g^{(1)}_*\varepsilon + \left(g^{(2)}_* + g^{(3)}_* \right )\varepsilon^2$ and after Taylor
expansion of $g_*^{1/3}$ in (\ref{nue}) we obtain:
\begin{equation}
\bar \nu \approx \nu_*
\left(\frac{D_0}{g_*^{(1)}\varepsilon}\right)^{1/3}
k^{-\frac{2\epsilon + \eta}{3}}\,,\qquad \nu_*\equiv
1-\frac{(g_*^{(2)}+ g_*^{(3)})\varepsilon}{3 g_*^{(1)}}\,. \label{nue1}
\end{equation}
%
Remind that for Kolmogorov values $\varepsilon =\eta=4/3$ the
exponent in (\ref{nue1}) becomes equal to $-4/3.$ Let us estimate
the contribution of helicity to the effective diffusivity in
the fixed point (\ref{fix}). In this point
$\varepsilon=\eta$ $((2\varepsilon+\eta)/3=\varepsilon)$ and
\begin{eqnarray}
\nu_* &=& 1- \varepsilon
\bigg[(\frac{(3+u_*)}{30(1+u_*)^3}\,\cdot\,{_2F_1}\left(1,1;\frac{7}{2};\frac{1}{(1+u_*)^2}\right)
\nonumber\\
 & - & \frac {\pi \rho^2}{12(1+u_*)}\,\cdot\,
{_2F_1}\left(\frac12,\frac12;\frac{5}{2};\frac{1}{(1+u_*)^2}\right)\bigg].\label{nukon}
\end{eqnarray}

\input epsf
\begin{figure}[t]
    \begin{flushleft} \leavevmode
    \epsfxsize=6.5cm
    \epsffile{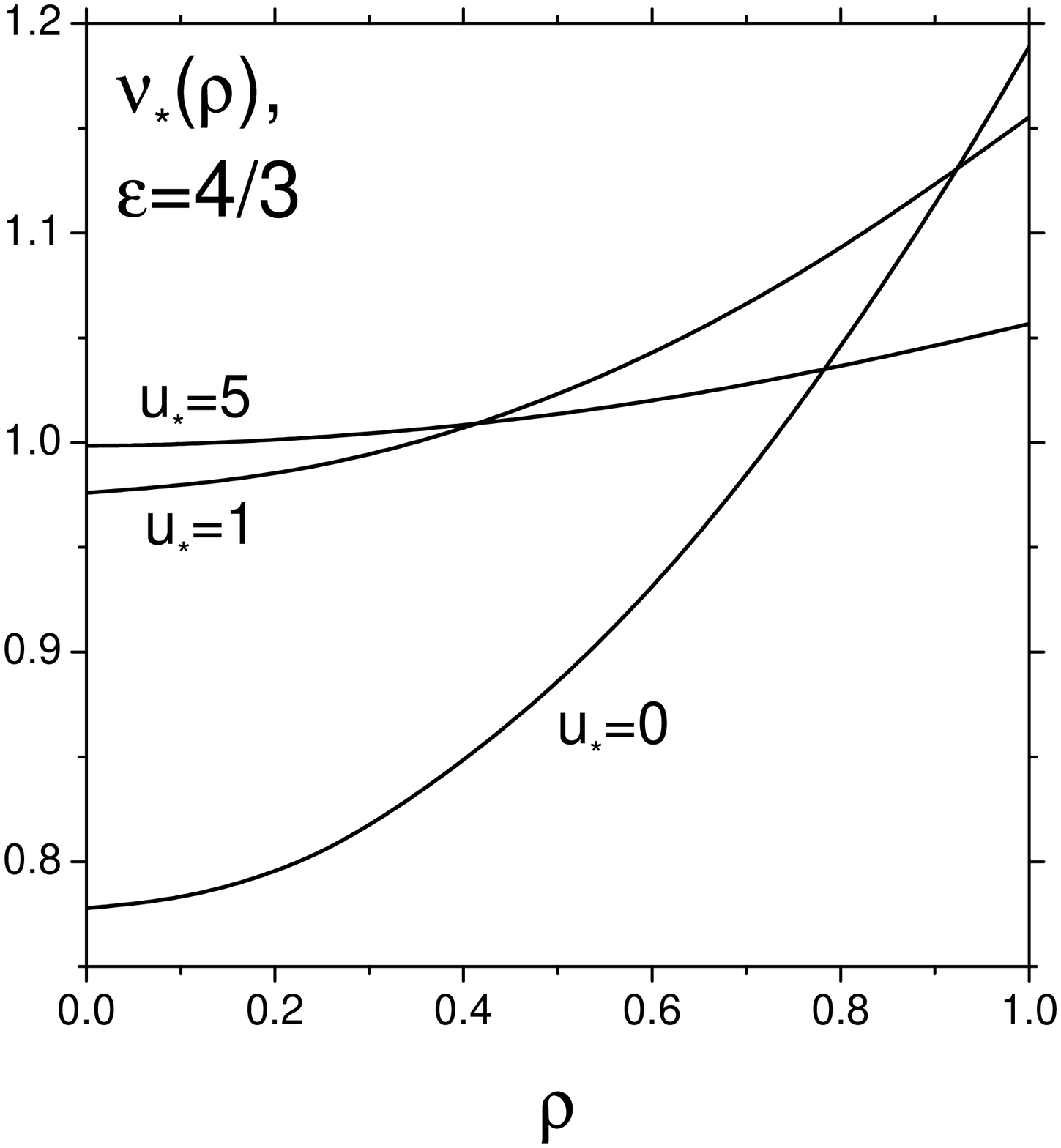}
  \end{flushleft}
\vspace{-10.1cm}
 \begin{flushright}
    \leavevmode
    \epsfxsize=6.5cm
    \epsffile{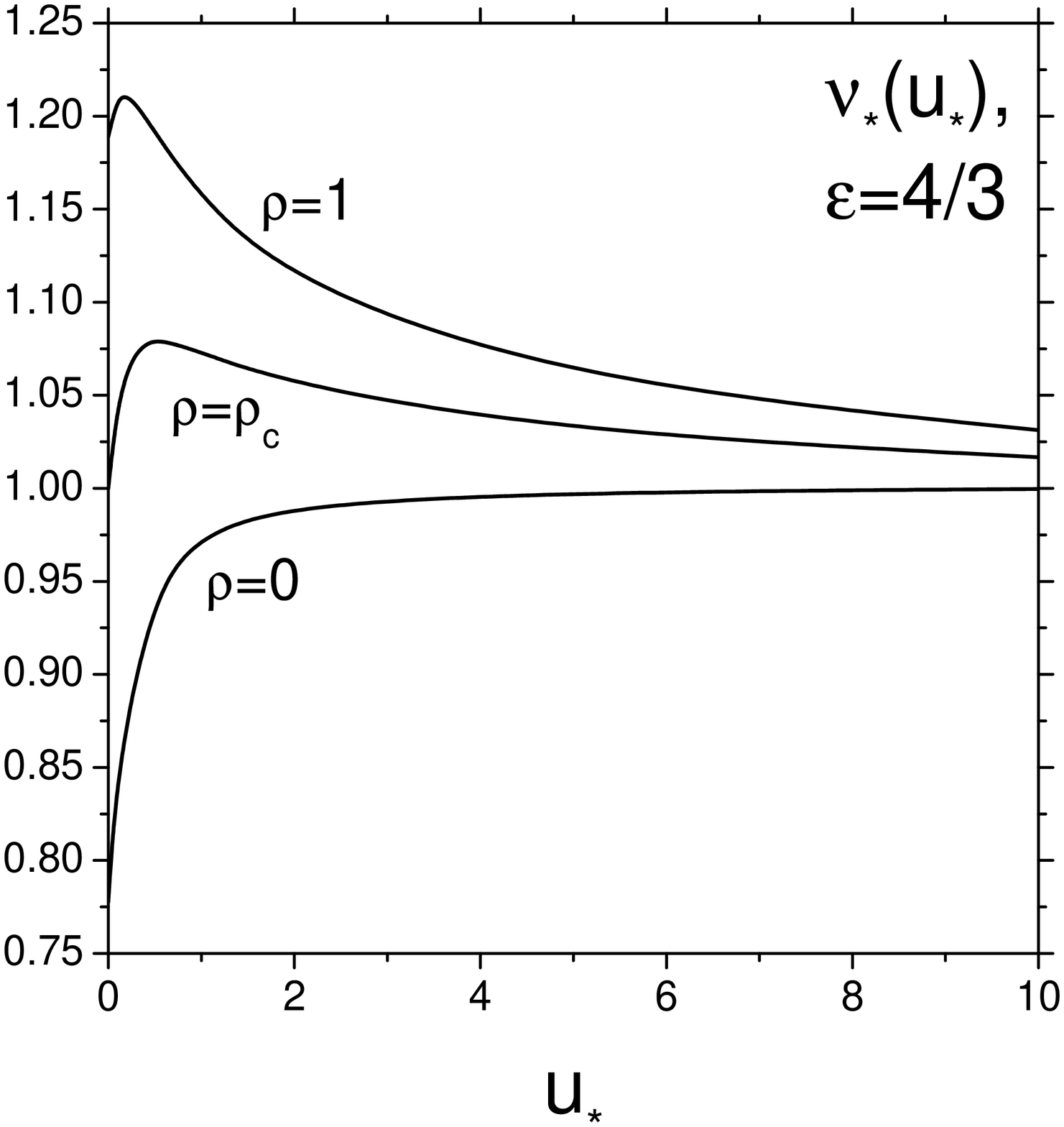}
\end{flushright}
\vspace{-1.5cm}\caption{(Left) The dependence of $\nu_*$ on the
helicity parameter $\rho$ for definite IR fixed point values $u_*$
of the parameter $u$. (Right) The dependence of $\nu_*$ on the IR
fixed point $u_*$ for the concrete values of the helicity parameter
$\rho$. The value $\rho_c=4/\sqrt{3}$. It is a special value related
to the analysis of the stability of the scaling regime which is not
discussed here. \label{fig5}}
\end{figure}

In figure \ref{fig5}, the dependence of the $\nu_*$ on the helicity
parameter $\rho$ and the IR fixed point $u_*$ for Kolmogorov value
of parameter $\varepsilon$ is shown . As one can see from these
figures when $u_*\rightarrow\infty$ (the rapid change model limit)
the two-loop corrections to $\nu_*=1$ are vanishing. Such behavior
is related to the fact that within the rapid change model there are
no two and higher loop corrections at all. On the other hand, the
largest two-loop corrections to the $\nu_*$ are given in the frozen
velocity field limit ($\nu_*\rightarrow0$).

Finally, let us analyze  time-behaviour of retarded response function
$G\equiv<\theta\theta'>$  in the limit $t \to \infty$.

In  frequency-wave vector representation $G(\omega, {\mathbf p})$ satisfies  Dyson equation (\ref{Dyson}).
 Self-energy operator $\Sigma$  is expressed via multi-loop Feynman graps and
 can be calculated perturbatively.
 We have found its divergent part up to two-loop approximation and calculated
 its finite part  with  the one-loop precision.

Using Dyson equation we find response function in time-wave vector representation:
\begin{equation}
G(t,\mathbf{p})=\int \frac{\mathrm{d}\omega}{2\pi} e^{-i\omega t}G(\omega,\mathbf{p}) =
\int \frac{\mathrm{d}\omega}{2\pi}\frac{e^{-i\omega t}}{-i\omega+\nu_{0}p^{2}-\Sigma(\omega,\mathbf{p})}\, .
\label{eq:Dyson1}
\end{equation}

In the lowest approximation  $\Sigma(\omega,\mathbf{p}) = 0$,  thus
the integral can be easily calculated:
$G_{0}(t,\mathbf{p})=\theta(t) \exp^{-i\omega_r t}.$ Here
$\theta(t)$ denotes usual step function and $\omega_r$ is residuum
in complex plain $ \omega$ in point $-i\nu_0 p^2 $. According to
\cite{Loran}, where analogical problem have been analyzed for turbulent viscosity,  
we suppose that this situation remains the same for
the full response function $G$, i.e., the leading contribution to
its  asymptotic behavior for $t \to \infty$ is determined by the
residuum $\omega =\omega_r$, which corresponds to the smallest root
of dispersion relation
\begin{equation}
G^{-1}(\omega,\mathbf{p})=-i\omega_r+\nu_{0}p^{2}-\Sigma(\omega_r,\mathbf{p})=0.
\label{eq:Dyson2}
\end{equation}
It is advantageous  to rewrite  the last relation in dimensionless
form:
\begin{equation}
1-z - I(1,z)= 0\,,\quad  z\equiv \frac{i\omega_r}{\nu_0 p^2} \,, \quad  I(1,z,g)\equiv \frac{ \Sigma(\omega,\mathbf{p})}{\nu_0 p^2}\,,
\end{equation}
which after renormalization can be rewritten in the fixed point $g_*$ (\ref{fix}) as follows
\begin{equation}
1-z_* - I_{*}= 0\, , \qquad \quad  z_*\equiv \frac{i\omega_r}{\bar{\nu} p^2} \,,
\label{ura}
\end{equation}
where $\bar{\nu}$ is effective diffusivity (\ref{nue1}) and
$I_*\equiv I_{*}(1,z_*, g_*)$ is renormalized (finite) part of
dimensionless self-energy operator $I$ at the fixed point $g_*$.

Hence decay law $G_{0}(t,\mathbf{p})\sim \exp^{-\nu_0 p^2 t}$  is changed into
\begin{equation}
G(t,\mathbf{p}) \sim  \exp^{-i\omega_r t}= \exp^{-z_*\bar{\nu} p^2 t}\, \qquad  t \to \infty \,.
\label{inf}
\end{equation}
To find the residuum $\omega_r$ (or, equivalently, $z_*$) it is necessary to calculalate quantity $I_*$. In
one-loop (linear in $\varepsilon$) approximation it can be written in the form:


\begin{equation}
I_*=-g_*\int_{-1}^{1}(1-x^{2})^{\frac{d-1}{2}}\mathrm{d}x\,{\cal I}\,
\label{eq:B.I}\end{equation}
with
\begin{equation}
{\cal I}=\int_{0}^{\infty}\mathrm{d}k\left[\frac{k}{ 1- z_*+(1+u_*)k^{2}-2kx}-\frac{\theta(k-1)}{(1+u_*)k}\right]\,.
\label{integral}
\end{equation}
Generally, the root $z_*$ can be complex and in one-loop  approximation it has form
\begin{equation}
z^{*}  =  z_{1}^{*}+\mathrm{i}z_{2},\quad  z_{1}^{*}  =  1+x_{1}\varepsilon,\quad
z_{2}^{*}  =  x_{2}\varepsilon\,. \label{eq:B.zp}\end{equation}

With our guaranteed precission 
$I_*$ is linear in $\varepsilon$, therefore on the first sight  it
seems that in the last integral it is enought to take $z_*=1$ ($g_*
\sim \varepsilon$), but, actually, for its correct calculation  we
need to remain imaginary part $ x_2\varepsilon.$ Then the integral
(\ref{integral}) can be easily calculated by means of Sokhotsky's
formula:
\begin{equation}
\lim_{\varepsilon\rightarrow 0^{+}}\frac{1}{y\pm\mathrm{i}\epsilon}=\mp\mathrm{i}\pi\delta(y)+P(\frac{1}{y}),
\end{equation}
where
$P(\frac{1}{y})$ denotes the principal value of the integral.
Integration over angle $x$
gives final result for dimensionless self-energy operator $I_*$:
\begin{eqnarray}
I_*  =  -\frac{g_*}{(1+u_*)}\left(\frac{\sqrt{\pi}\Gamma(d+1)
\left(\gamma+\psi(1+\frac{d}{2})+2\ln|1+u_*|\right)}{d\,2^{d}\Gamma\left(\frac{d-1}{2}\right)\Gamma\left
(\frac{d}{2}+1\right)}\pm\mathrm{i}\pi\frac{d-1}{2d}\right),\nonumber \\
\label{eq:B.vysledok}\end{eqnarray}
where $\gamma$ is Euler's constant
and $\psi(z)$ is digamma function defined as $\psi(z)=\Gamma^{\prime}(z)/\Gamma(z)$.

Successful calculation of integral $I$ allows one to determine  the residue
$z_*$ (\ref{eq:B.zp}).
Comparison of real and complex part of both sides of (\ref{ura})
gives the following terms in real space $d=3$ and in the fixed point  $g_*=3(1+u_*)/2$
(see (\ref{fix}))
\begin{equation}
x_{1}  = \frac{8}{3}+2\ln\frac{1+u_*}{2},\qquad
x_{2} =  \pm \frac{\pi}{2}.
\end{equation}

Due to the existence of two complex conjugate values $z_*$  the response function $G(t,p^{2})$ can be written in the asymptotic limit $t\rightarrow\infty$ in the following final form
\begin{equation}
G(t,p^{2})\cong e^{-\nu_{eff} p^{2-\varepsilon}t}\sin(\nu_f p^{2-\varepsilon}t),
\end{equation}
where
\begin{eqnarray}
 \nu_{eff}&\equiv& \bigg[1- \varepsilon \bigg[ \frac{8}{3}+2\ln\frac{1+u_*}{2}+
\frac{(3+u_*)}{30(1+u_*)^3}\,\cdot\,{_2F_1}\left(1,1;\frac{7}{2};\frac{1}{(1+u_*)^2}\right)
\nonumber\\
 & - & \frac {\pi \rho^2}{12(1+u_*)}\,\cdot\,
{_2F_1}\left(\frac12,\frac12;\frac{5}{2};\frac{1}{(1+u_*)^2}\right)\bigg]\bigg]
 \left(\frac{2 D_0}{3(1+u_*)\varepsilon}\right)^{1/3}\\
 \nu_f &\equiv& \frac{\pi\varepsilon}{2} \left(\frac{2 D_0}{3(1+u_*)\varepsilon}\right)^{1/3}
\nonumber
\end{eqnarray}
It's clear that the exponential
damping is accompanied by the oscilations.

\section{Conclusion}
We have studied  the advection of scalar field by turbulent flow in the frame of extended Kraichnan model and
investigated the influence of helicity  on anomalous scaling, stability of asymptotic regimes and effective
diffusivity. Such investigation is useful for understanding of efficiency of simplified  models
to study  the real turbulent motions by means of modern theoretical methods including renormalization group approach.
Actually, we performed two-loop calculations of divergent parts of Feynman graphs, which are necessary to achieve
multiplicative renormalization of equivalent field theoretic model. We have shown that anomalous scaling, which
is typical for the Kraichnan model and its numerous extensions \cite{Juha2,Pressure}, is not violated by inclusion of helicity to
the incompressible fluid.
On the other hand, stability of asymptotic regimes, values of fixed RG points and turbulent diffusivity strongly depend on amount of helicity.
It can be easily see  from (\ref{nukon}) that helicity enlarges turbulent diffusivity and high order contributions lead to the
appearance of oscillations in response function (\ref{eq:Dyson1}).

\section{Acknowledgement}
M.H. is thankful to N.V.Antonov and L.Ts. Adzhemyan for discussion.
The work was supported in part by VEGA grants 3211 and 2/6193/26 of
Slovak Academy of Sciences, by Science and Technology Assistance
Agency under contract No. APVT-51-027904, by  grant RFFI - RFBR
05-05-64735, by  grant RFFI - RFBR 05-02-17603, and by COFIN 2003
"Sistemi Complessi e Problemi a Molti Corpi.".


\section*{References}
\begin{thebibliography}{99}

\bibitem{Monin} Monin A S and  Yaglom A.M. 1975 {\it Statistical Fluid
Mechanics} Vol.2 (Cambridge: MA MIT Press)



\bibitem{McComb}  McComb W D  1990 {\it The Physics of Fluid Turbulence}
(Oxford: Clarendon)

\bibitem{Legacy}  Frisch U 1995   {\it Turbulence: The Legacy of A.~N.~Kolmogorov}
(Cambridge: Cambridge University Press)

\bibitem{Dyn} Bohr T,  Jensen M H,  Paladin G and   Vulpiani A 1998
{\it Dynamical Systems Approach to Turbulence}
(Cambridge: Cambridge University Press)

\bibitem{Burgulence} Bec J and Frisch U 2000 Burgulence
{\it Preprint} nlin.CD/0012033

\bibitem{FGV} Falkovich G, Gaw\c{e}dzki K and Vergassola M 2001
{\it Rev. Mod. Phys.} {\bf 73} 913


\bibitem{Obu49}
Obukhov A M 1949 {\it Izvestiya Akademii Nauk SSSR: Geografia Geofizika} {\bf 13} 58

\bibitem{Kra68}
Kraichnan R H 1968 {\it Phys. Fluids} {\bf 11} 945

\bibitem{AnHoGaAn84}
Antonia R A, Hopfinger~E~J, Gagne~Y and Anselmet~F 1984 {\it Phys.
Rev.} A {\bf 30}, 2704

\bibitem{Sre91}
Sreenivasan~K~R 1991 {\it Proc. R. Soc. London} Ser. A {\bf 434} 165
\bibitem{HolSig94}
Holzer~M and Siggia~E~D 1994 {\it Phys. Fluids} {\bf 6}, 1820
\bibitem{Kraich1} Kraichnan R H 1994 {\it Phys. Rev. Lett.} {\bf 72} 1016



\bibitem{GK}
Gaw\c{e}dzki K and  Kupiainen A 1995 {\it Phys. Rev. Lett.} {\bf 75} 3834



\bibitem{Pumir}
Pumir~A 1997 {\it Europhys. Lett.} {\bf 37} 529
Pumir~A 1998 {\it Phys. Rev.} E {\bf 57}, 2914
\bibitem{AveMaj90}
Majda A 1993 {\it J. Stat. Phys.} {\bf 73} 515

\bibitem{Zinn} Zinn-Justin J 1989  {\it Quantum Field Theory and
Critical Phenomena} (Oxford:  Clarendon)

\bibitem{book} Vasil'ev A N 2004  {\it Quantum-Field Renormalization
Group in the Theory of Critical Phenomena and Stochastic Dynamics}
(New York: Chapman\& Hall/CRL Press)

\bibitem{RG} Adzhemyan L Ts, Antonov N V and Vasil'ev A N 1998
{\it Phys. Rev.} E {\bf 58} 1823



\bibitem{turbo} Adzhemyan L Ts Antonov N V and Vasil'ev A N 1999
{\it The Field Theoretic Renormalization Group in Fully Developed
Turbulence} (London: Gordon \& Breach)

\bibitem{RG1} Adzhemyan L Ts, Antonov N V,  Barinov V A,
Kabrits Yu S and Vasil'ev A N  2001 {\it Phys. Rev.} E {\bf 63}
025303(R);  2001 {\it Phys. Rev.} E {\bf 64} 019901(E); 2001 {\it
Phys. Rev.} E {\bf 64} 056306

\bibitem{Novikov} Adzhemyan L Ts, Antonov N V,  Hnatich M and Novikov S
2000 {\it Phys. Rev.} E {\bf 63}  016309

\bibitem{RG3} Antonov N V 1999 {\it Phys. Rev.} E {\bf 60} 6691
\item[] Antonov N V  2000 Physica D {\bf 144} 370

\bibitem{Lanotte2} Antonov N V, Lanotte A and Mazzino A 2000
{\it Phys. Rev.} E {\bf 61} 6586

\bibitem{amodel} Adzhemyan L Ts, Antonov N V, Mazzino A,
Muratore-Ginanneschi P and  Runov A V  2001 {\it Europhys. Lett.}
{\bf 55} 801

\bibitem{all2}
Shraiman B I and Siggia E D 1994 {\it Phys. Rev.} E {\bf 49} 2912
\item[] Shraiman B I and Siggia E D 1996 {\it Phys. Rev. Lett} {\bf 77}  2463

\bibitem{Juha2} Adzhemyan L Ts, Antonov N V and Honkonen J 2002
{\it Phys. Rev.} E {\bf 66}  036313


\bibitem{dynamo1} Vainstein S I, Zel'dovich Ya A and Ruzmaykin A A 1980
{\it  Turbulent
dynamo in astrophysics} (Moskva: Nauka) (in Russian)

\bibitem{Moffatt}Moffatt H K 1978 {\it Magnetic field generation in electrically conducting fluids}  (Cambridge: Cambridge univ. press)







\bibitem{Etling85} Etling D 1985 {\it Beitr. Phys. Atmosph.}  {\bf 58}  88



\bibitem{Ponomar2003}
Moffat K H and Tsinober A 1992 {\it Annu. Rev. Fluid Mech.} {\bf 24} 281 
\item[] Ponomarev~V~M, Khapaev~A~A and Chkhetiani~O~G 2003
{\it  Izvestya: atmospheric and oceanic physics} {\bf 39(4)} 391


\bibitem{Brissaud73}Brissaud~A, Frisch~U, Leorat~J, Lesieur~M and Mazure~A
1973  {\it Phys. Fluids} {\bf 16} 1363

\bibitem{Moiseev1996} Moiseev~S~S and Chkhetiani~O~G 1996
 {\it JETP} {\bf 83}  192

\bibitem{koprovall}
Koprov B M, Koprov V M, Ponomarev V V,Chkhetiani O G 2005
{\it Doklady Physics} {\bf 50} 419  

\bibitem{Chkhet96} Chkhetiani~O~G 1996  {\it JETP Letters} {\bf 63}  808

\bibitem{kurien}
Kurien S,  Taylor M A,  Matsumoto T 2004  
{\it Journal of Fluid Mech.} {\bf 515} 87 

\bibitem{Kraich73} Kraichnan~R~H 1973 {\it J. Fluid Mech.} {\bf 59} 745



\bibitem{Chen2003} Chen~Q, Chen~S and Eyink~G~L 2003 {\it Phys. Fluids} {\bf 15} 361

\bibitem{Pouquet78}
Pouquet A, Fournier~J~D and Sulem~P~L 1978  {\it J. De Physique Lett.}
 {\bf 39(13)} 199

\bibitem{Zhou91}  Zhou Ye 1991 {\it Phys. Rev.} A. {\bf 41}  5683


\bibitem{Drum84} Drummond~S~T, Duane~S and  Horgan~R~P 1984
{\it J.Fluid Mech.} \textbf{138}  75

\bibitem{Knobl77}  Knobloch~E 1977
{\it J. Fluid Mech.} \textbf{83}  129

\bibitem{Lipscombe91} Lipscombe T C, Frencel~A~L and ter Haar~D 1991
{\it J. of Stat.Phys.} \textbf{63}  305


\bibitem{Belian1998} Belyan A V, Moiseev S S, Golbraih~E~I and
Chkhetiani~O~G 1998
{\it Physica} A {\bf 258} 55

\bibitem{Dolg87} Dolginov A Z and Silantiev~N~A 1987
{\it JETP} {\bf 93} 159

\bibitem{Drummond2001} Dean~D~S, Drummond~I~T and Horgan~R~P 2002
{\it Phys. Rev. E}  {\bf 63}  61205
%
%
%

\bibitem{Eyink96}
Eyink G 1996 {\it Phys. Rev.} E {\bf 54} 1497

\bibitem{Bouchaud}
Bouchaud J P and Georges A 1990 {\it Phys. Rep.} {\bf 195} 127
\bibitem{Honkonen}
Honkonen J, Pis'mak Yu M and Vasil'ev A N  1989 {\it J. Phys.} A  {\bf 21} 835

\bibitem{Martin}Martin P C, Siggia E D and Rose H A 1973
{\it Phys. Rev.} A {\bf 8} 423

\bibitem{Loran}
Adzhemyan L V  and  Adzhemyan L Ts  {\it Vestnik Sankt Peterburgskoho Universiteta
2003 Vestnik Sankt Peterburgskoho Universiteta:
Seria Fizika Chemia} {\bf 4(28)} 94 (in Russian)


\bibitem{Pressure} Antonov N V, Hnatich M, Honkonen J and Jurcisin M 2003
{\it  Phys. Rev.} E {\bf 68} 046306

\end{thebibliography}
\end{document}